# Do the Kepler AGN Light Curves Need Re-processing?

Vishal P. Kasliwal,[1]* Michael S. Vogeley,[1] Gordon T. Richards,[1] Joshua Williams,[2] and Michael T. Carini[2]
[1]*Department of Physics, Drexel University, Philadelphia, PA 19104-2875, USA*
[2]*Department of Physics and Astronomy, Western Kentucky University, Bowling Green, KY 42101-1077, USA*

**ABSTRACT**
We gauge the impact of spacecraft-induced effects on the inferred variability properties of the light curve of the Seyfert 1 AGN Zw 229-15 observed by Kepler. We compare the light curve of Zw 229-15 obtained from the Kepler MAST database with a re-processed light curve constructed from raw pixel data (Williams & Carini 2015). We use the first-order structure function, $SF(\delta t)$, to fit both light curves to the damped power-law PSD of Kasliwal, Vogeley & Richards (2015). On short timescales, we find a steeper log-PSD slope ($\gamma$ = 2.90 to within 10 percent) for the re-processed light curve as compared to the light curve found on MAST ($\gamma$ = 2.65 to within 10 percent)—both inconsistent with a damped random walk which requires $\gamma$ = 2. The log-PSD slope inferred for the re-processed light curve is consistent with previous results (Carini & Ryle 2012; Williams & Carini 2015) that study the same re-processed light curve. The turnover timescale is almost identical for both light curves (27.1 and 27.5 d for the reprocessed and MAST database light curves). Based on the obvious visual difference between the two versions of the light curve and on the PSD model fits, we conclude that there remain significant levels of spacecraft-induced effects in the standard pipeline reduction of the Kepler data. Re-processing the light curves will change the model inferenced from the data but is unlikely to change the overall scientific conclusion reached by Kasliwal et al. (2015)—not all AGN light curves are consistent with the DRW.

## 1 INTRODUCTION

The underlying causes of the large, stochastic variations seen in active galactic nuclei (AGN) flux across the electromagnetic spectrum are not well understood (Peterson 1997). Continuum variations in the optical arise in the accretion disk of the AGN and hence may be caused by inward propagating stochastic viscosity fluctuations (Lyubarskii 1997; Cowperthwaite & Reynolds 2014). Alternatively, X-ray flares may be the drivers of optical variability (Goosmann et al. 2006). Observations of quasar variability in ground based surveys such as the PG (Kelly et al. 2009), OGLE (Kozłowski et al. 2010), and the SDSS (MacLeod et al. 2010) suggest that the observed light curves are well described by a damped random walk (DRW) (Kelly et al. 2009). The PSDs of serendipitously observed AGN light curves from the NASA Kepler planet-finding mission have been found be significantly steeper than permitted by a DRW (Mushotzky et al. 2011; Carini & Ryle 2012; Edelson et al. 2013; Wehrle et al. 2013; Edelson et al. 2014; Revalski et al. 2014; Kasliwal et al. 2015). While the Kepler AGN light curves are unrivaled in precision, cadence and regularity, they may suffer from serious processing issues. We have found that the pipeline-supplied Kepler light curve of the Seyfert 1 galaxy Zw 229-15 is not consistent with ground-based photometric observations obtained over the duration of the Kepler mission. A re-processing of the light curve of Zw 229-15 using ground-based photometry to improve accuracy suggests that the light curve of this object remains inconsistent with the DRW.

## 2 THE Kepler MISSION

Launched in 2009 as a NASA *Discovery Mission*, the Kepler spacecraft has been designed to survey the Milky-Way for Earth-sized and smaller planets around the habitable zone by using the transit-method to detect exo-planets (Borucki et al. 2010). To meet the required photometric precision, Kepler was designed as a 0.95 m Schmidt camera with a

* E-mail: vpk24@drexel.edu

115.6 deg$^2$ field of view (FOV). The imager consists of 42 CCDs of size 50×25 mm with 2200×1024 pixels each, resulting in an image scale of 3.98 arcsec per pixel. The CCDs are arranged in 21 modules with 4 readout channels per module. Each integration consists of a 6.02 s exposure followed by a 0.502 s readout. The long-cadence (LC) light curves of interest for AGN variability studies are created by combining 270 integrations resulting in a sampling interval of 29.43 min (Van Cleve & Caldwell 2009; Christiansen et al. 2013). Precision photometry requires that no pixel contains more than 60 per cent of the flux from a point source target (Gilliland 2004)—hence the Kepler PSF is large (95 % encircled energy in a diameter of 6.4 pixels).

Photometry from Kepler is of very high precision (1 part in $10^5$), but suffers in accuracy. Raw light curves extracted using simple aperture photometry (SAP) are contaminated by thermally-driven focus variations, pointing offsets, and differential velocity aberration. Some of the readout channels on certain CCDs have performance issues such as out-of-spec read noise levels and gain. Moire and rolling band effects are present in some channels (Kolodziejczak et al. 2010). Most of these issues are tracked in the form of Co-trending Basis Vectors (CBV) that quantify effects common to multiple targets across the FOV. The CBVs can be used to mostly eliminate spacecraft-induced trends in the dataset (Stumpe et al. 2012; Smith et al. 2012), however they can also remove real variability in the light curves of AGN.

More troubling is the crowding caused by the (intentionally) large Kepler point spread function (PSF). The large PSF can cause significant contamination from nearby sources when performing simple aperture photometry. The Kepler Science Operations Center (SOC) pipeline determines the photometric aperture based on target brightness. As a consequence, the light curves of bright extended targets can have significant levels of contamination from surrounding objects.

## 3 THE LIGHT CURVES

We assess the impact of crowding and CBV induced feature removal on the photometry of AGN targets by comparing the Kepler SOC pipeline-generated light curve of Zw 229-15 from Data Release 23 (DR23) available in the Mikulski Archive for Space Telescopes (MAST) database (Thompson et al. 2013)—hereafter the *MAST_DR23* light curve—to the re-processed version of Williams & Carini (2015)—hereafter the *CW2015* light curve—that uses a smaller aperture mask to extract the flux of the AGN at the center of Zw 229-15 and is verified by ground based photometry.

### 3.1 The *MAST_DR23* Light Curve

For the pipeline-processed light curve, we use the DR23 light curve available in the MAST database. *MAST_DR23* was processed using version 9.1.4 of the Kepler SOC pipeline (Fanelli et al. 2011). Version 9.1.4 of the Kepler pipeline uses the Presearch Data Conditioning-Maximum A Posteriori (PDC-MAP) algorithm of Stumpe et al. (2012) and Smith et al. (2012) to remove spacecraft-induced systematics from light curves. This goal of the PDC-MAP as implemented in the PDC module—to effectively remove spacecraft-induced variability while leaving variability intrinsic to the source intact—is achieved by using a Bayesian prior to help determine the level of spacecraft induced contamination.

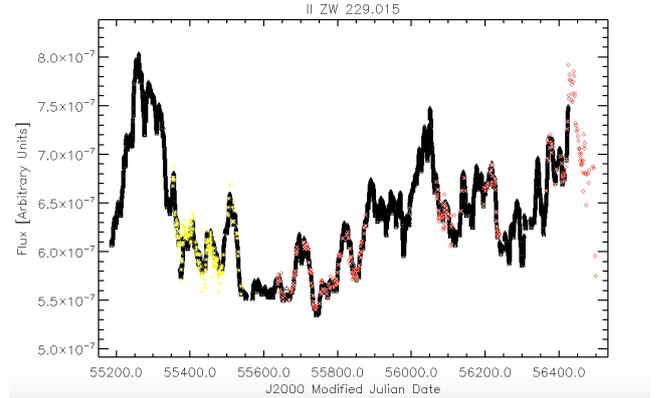

**Figure 1.** The *CW2015* re-processed light curve of Zw 229-15. The black points are the re-processed Kepler observations. The yellow points are ground-based observations obtained as part of the Barth et al. (2011) reverberation mapping campaign while the red points are observations made with the RCT (Williams & Carini 2015; Gelderman et al. 2003) by the WKU group.

The operation of the PDC module is described in Smith et al. (2012). In brief, cross-correlation is used to select quiet targets in the Kepler FOV—the light curves of targets that are intrinsically non-variable should be highly correlated since the only signal comes from the spacecraft-induced systematics. The quiet targets are used to construct 16 CBVs by performing a singular value decomposition analysis. These CBVs capture all the features introduced into the light curves of Kepler targets by spacecraft systematics. Linear least-squares may then be used to determine the weights to assign to combinations of these CBVs when de-trending AGN light curves. The drawback of this procedure is that any co-incidence between the intrinsic variability features in the light curve of interest and the CBVs will be projected out of the final light curve. The PDC-MAP algorithm avoids this phenomenon by using adjacent (in magnitude and position) targets to construct priors for the CBV weights. By biasing the CBV weights towards the values most likely to remove spacecraft trends in nearby objects, the PDC-MAP algorithm is able to optimally remove spacecraft trends while leaving the intrinsic variability un-altered.

We are aware of a few caveats for using the CBVs to correct for instrument noise. At the photometric precision level of Kepler, 'quiet' stars do not exist (Gilliland et al. 2011). The inflight photometric precision for a $m_{Kepler}$ = 12 star exceeds the budgeted precision level of 20 ppm by ∼ 75 per cent (Thompson et al. 2013) depending on the quarter under consideration. As pointed out in Gilliland et al. (2011), this is primarily caused by intrinsic stellar variability and is unavoidable. Therefore, the CBVs are constructed using the *quietest* stars rather than quiet stars and are contaminated by trace amounts of genuine astrophysical variability. Consequently, the true astrophysical variability of the so-called quiet sources acts as a source of noise. Another concern is that the entire approach of the CBVs is incorrect for studying astrophysical variability even if it is sufficient for planetary detection: the CBVs are applied as a *subtractive* correction to remove what are primarily *multiplicative* systematics.

Finally, the *MAST_DR23* light curve exhibits large discontinuities in flux between quarters. This phenomenon is caused by the re-definition of the photometric target aperture between quarters and is entirely non-physical. Following Kasliwal et al. (2015), we perform 'stitching' to remove these inter-quarter discontinuities by matching the average flux in the 96 observations at the endpoints of quarters. The stitching technique used by us is based on suggestions in Kinemuchi et al. (2012) and is similar to the technique used by Revalski et al. (2014).

### 3.2 The Re-Processed *CW2015* Light Curve

The re-processed *CW2015* light curve was created by re-defining the target aperture using the Python tool KEPEXTRACT to match ground-based observations as described in Carini & Ryle (2012) and in Williams & Carini (2015). The Western Kentucky University (WKU) ground-based observations were obtained with the 1.3 m Robotically Controlled Telescope (RCT) (Gelderman et al. 2003). Observations were obtained through a Johnson V filter using a cryogenically cooled CCD camera system with a SiTE CCD. Magnitudes were determined using differential photometry techniques employing the comparison sequence found in Barth et al. (2011). Figure 1 shows the *CW2015* light curve along with the ground-based photometric measurements and error bars (smaller than the plotting symbols). Optimal photometric apertures for Kepler targets are determined based on the target luminosity. The 14 pixel pipeline aperture for Zw 229-15 covers 221.8 arcsec$^2$ and includes significant contamination from the host galaxy. The reverberation mapping campaign of Barth et al. (2011) suggests that the correct photometric aperture is 50.24 arcsec$^2$. Host galaxy contamination was minimized by constructing the *CW2015* lightcurve using a 2-pixel aperture (31.6 arcsec$^2$). This aperture was found to minimize the difference between the Kepler light curve and the ground based observations while simultaneously removing the need to perform the 'stitching' required in the case of the *MAST_DR23* light curve. As is evident from the close correspondence between the ground-based photometry and the *CW2015* light curve in figure 1, optimal re-extraction of the flux of Zw 229-15 has been achieved.

### 3.3 Comparison of the *MAST_DR23* and *CW2015* Light Curves

Figure 2 shows the *MAST_DR23* (orange) and *CW2015* (purple) light curves of Zw 229-15. The *MAST_DR23* light curve is scaled to the *CW2015* light curve for the purpose of illustration—the scaling is not used in the actual analysis. Visually, the two versions of this light curves exhibit significant differences, especially during the first 700 d. A close examination reveals that although the two light curves are not co-incident, the same 'features' appear in both light curves. The same peaks and valleys are present, but their relative sizes are different in the *CW2015* light curve as compared to the *MAST_DR23* light curve i.e. the (time-dependent) gains applied by the two processing strategies are different. We use the damped power law (DPL) model of Kasliwal et al. (2015) to examine the difference between the two light curves. We expect that the inferred values of the DPL model parameters will be similar but not identical for these two light curves.

## 4 THE DAMPED POWER LAW PSD MODEL

The correlation structure of an AGN light curve is indicative of the physical processes responsible for the observed variability. A DRW light curve may be the result of accretion disk with local MHD hot-spots that appear at random and dissipate over some characteristic physical timescale (Dexter & Agol 2011). Correlations exist between succesive flux measurements in the DRW because the MHD hot-spots do not dissipate instantaneously. The DRW specifies the correlation structure of the light curve—the *Auto-Covariance Function* (ACVF) of the DRW is of the form

$$ACVF_{DRW}(\Delta t) = \sigma^2 e^{-\frac{\Delta t}{\tau}}, \qquad (1)$$

where $\tau$ is the decorrelation timescale and $\sigma^2$ is the amplitude of the ACVF on short timescales. The corresponding power spectral density (PSD) is

$$PSD_{DRW}(\nu) = \frac{4\sigma^2 \tau}{1 + (2\pi\nu\tau)^2}. \qquad (2)$$

Kasliwal et al. (2015) allow the correlation structure of the light curves to be different from that of the DRW by generalizing form of the DRW PSD in equation (2) to

$$PSD_{DPL}(\nu) = \frac{\sigma^2_{Eff}}{1 + (2\pi\nu\tau)^\gamma}, \qquad (3)$$

and gather the quantities in the numerator of equation (2) into a single variable $\sigma_{Eff}$. The parameter $\gamma$ is introduced to allow the logarithmic slope of the PSD to be a free parameter. If $\gamma = 2$, equation (3) recovers the DRW PSD of equation (2). $\gamma < 2$ implies that short timescale correlation structure is weaker than that of the DRW i.e. the time series looks less smooth as compared to the DRW. On the other hand, $\gamma > 2$ implies that the short-timescale correlation structure is stronger than that of the DRW i.e. the time series looks smoother than the DRW. The DPL model is a simplification of the popular bending power law (BPL) model of McHardy et al. (2004):

$$PSD_{BPL}(\nu) = A\nu^{-\alpha_1} \prod_{i=1}^{L} \frac{1}{1 + (\frac{\nu}{\nu_{Bi}})^{\alpha_{i+1}-\alpha_i}}, \qquad (4)$$

where $\alpha_i$ are the logarithmic PSD slopes above the corresponding bend frequencies $\nu_{Bi}$.

## 5 INFERENCE OF THE DPL MODEL PARAMETERS

We use structure functions to determine best-fit values of the parameter $\gamma$ in the DPL model of equation (3). We define the 1$^{st}$-Order Structure Function (SF) of the process $F(t)$ to be

$$SF(\delta t) = \langle [F(t + \delta t) - F(t)]^2 \rangle. \qquad (5)$$

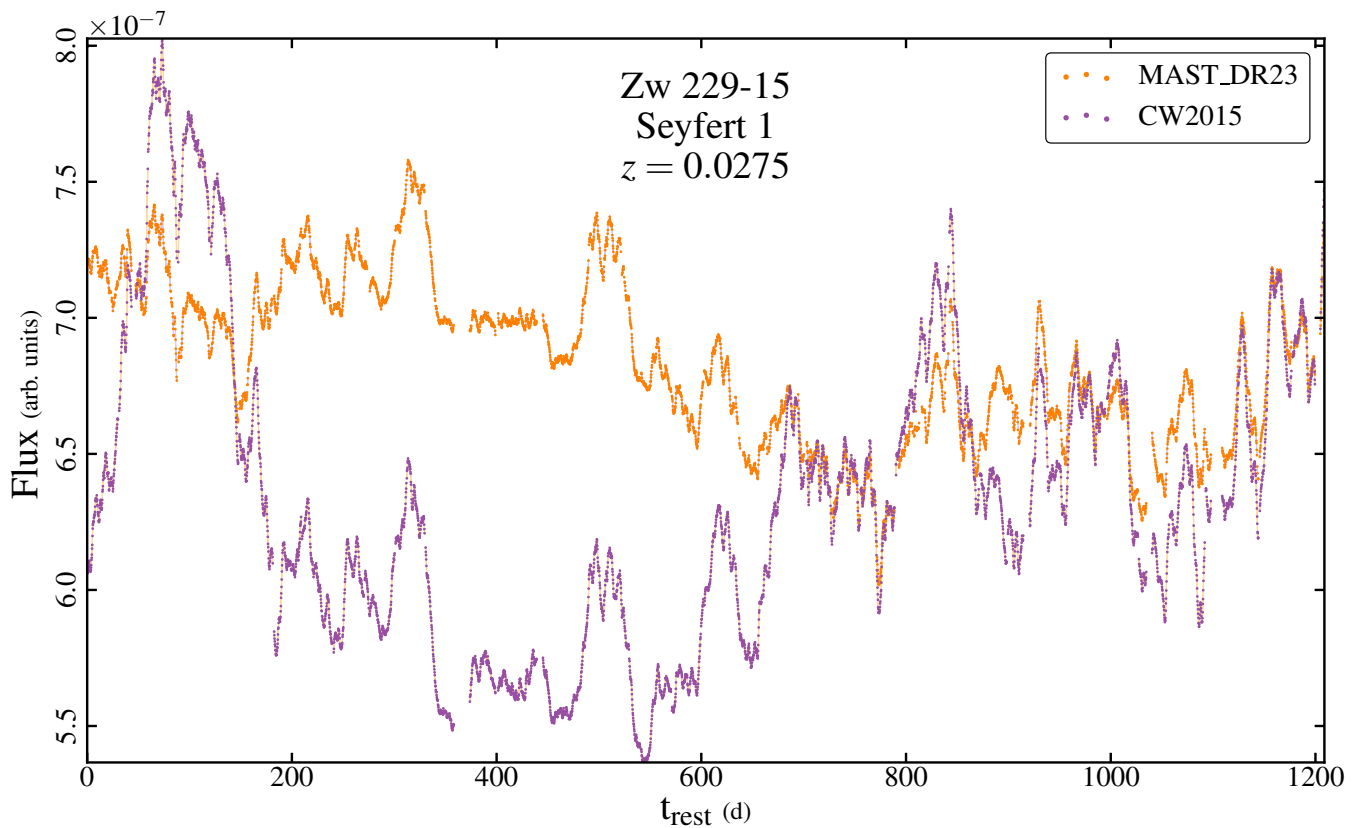

**Figure 2.** The *MAST_DR23* and *CW2015* light curves of Zw 229-15.

The structure function quantifies how the variance of the flux differences changes as a function of time-lag. Structure functions offer benefits over directly estimating the PSD and ACVF/ACF of the process. Unlike the PSD, structure functions may be estimated in real space as opposed to Fourier space, making them more robust estimators of model parameters than PSD estimators that suffer from windowing and aliasing concerns (Rutman 1978). Structure functions offer an advantage over estimating the ACVF and ACF because of the de-trending properties of structure functions: an $n^{\text{th}}$-Order Structure Function is insensitive to $(n-1)^{\text{th}}$-order trends in the dataset.

To compare the observed structure function of the Zw 229-15 light curves to those predicted by the DPL model, we use the Monte Carlo method of Kasliwal et al. (2015). Briefly, we use the light curve simulation method of Timmer & Koenig (1995) to generate mock light curves drawn from the DPL model. In order to decrease the impact of aliasing and include long timescale trends, we first create intermediate mock light curves of length $2^{23}$ observations (for optimal performance of the FFT routines) that are spaced at 1/10 the sampling interval of *Kepler* i.e. at ∼ 3 min. Intermediates created in this manner are 13.8 × longer (in duration) and oversampled 10 × as compared to the actual light curve of Zw 229-15. These intermediate mocks are subsequently down-sampled to match the sampling interval of the real light curve in the rest frame of Zw 229-15 ($\delta t = 28.64$ min at $z = 0.0275$). A segment of length equal to the actual length of the light curve is selected from the intermediate mock using a randomized starting location in order to include long timescale trends. The sampling pattern of the real light curve is matched exactly by substituting 0 into the mock light curves and creating a mask, $w_i$ to track the missing observations. The mocks generated in this manner are used to compute corresponding mock structure functions.

We compute the mean and variance of the $\log_{10} SF(\Delta T)$ (Kasliwal et al. 2015; Emmanoulopoulos et al. 2010) at each location in DPL model parameter space to compare the observed structure function of each version of the light curve of Zw 229-15. The non-linear optimizer COBYLA of Powell (1994) is used to compute the best-fit values of the DPL model parameters by minimizing

$$\chi^2 = \sum_{n=1}^{N-1} \frac{(\mu_n - \log_{10} SF_{n,obs})^2}{\sigma_n^2}, \quad (6)$$

where $\mu_n$ and $\sigma_n^2$ are the mean and variance of the ensemble of the $\log_{10} SF_n$ with $n = \Delta t/\delta t$ i.e. the $n^{\text{th}}$-lag. To ensure that the distribution of mock $\log SF$s is insensitive to the light curve realizations, we use $10^5$ mocks at every step in the optimization procedure.

Unfortunately the computing time required to estimate the value of $\chi^2$ at every location in DPL model parameter space makes it impossible for us to probe the full posterior distribution of the DPL model parameters to obtain confidence intervals on the parameter values. Instead, we generate short mock light curves with known input DPL model parameters and attempt to recover the input parameter val-ues using our estimation process. From these simulations, we estimate that the fractional uncertainties of our parameter values are ~ 10 percent.

## 6 OBSERVED STRUCTURE FUNCTIONS AND FITS

We fit both light curves of Zw 229-15 to the DPL PSD model. Figure 3 shows the observed structure functions corresponding to the *MAST_DR23* processing (orange) and the *CW2015* re-processing (purple). Also shown are the mean and standard deviations of the ensemble of structure functions corresponding to the best-fit DPL model parameters for the *MAST_DR23* light curve (red line + light-red shaded region) and *CW2015* re-calibrated light curve (blue line + light-blue shaded region). The structure function of the *MAST_DR23* light curve of the Zw 229-15 is scaled to match the structure function of the *CW2015* light curve—this scaling is purely for visualization and is not used in the actual analysis. Comparison of the goodness of fit for the *MAST_DR23* ($\chi^2_{DOF}$ = 1.43) and *CW2015* ($\chi^2_{DOF}$ = 0.71) indicates that while the *MAST_DR23* light curve contains complex behavior not captured by the DPL model, the *CW2015* light curve is somewhat overfit by the DPL model. This suggests that the *MAST_DR23* light curve contains information that the *CW2015* light curve does not—possibly the non-physical spacecraft-induced variability. The numerical value of the $\chi^2$ value of the *MAST_DR23* light curve falls in the 83$^{rd}$ percentile of the $\chi^2$ values observed in the Monte-Carlo simulated mock light curves. On the other hand, the $\chi^2$ value of the *CW2015* light curve falls in the 38$^{th}$ percentile of the Monte-Carlo simulated mock light curves suggesting that the DPL fit for the *CW2015* light curve is better than that for the *MAST_DR23* light curve.

The inferred value of the DPL model short-timescale log-PSD slope is steeper for the *CW2015* light curve ($\gamma$ = 2.90) as compared to the *MAST_DR23* light curve ($\gamma$ = 2.65) i.e. the *CW2015* light curve is smoother on very short timescales than the *MAST_DR23* light curve. Both light curves remain inconsistent with a damped random walk ($\gamma$ = 2) suggesting that it is unlikely that the non-DRW behavior observed in Kepler AGN light curves is attributable to data-quality issues. The de-correlation timescale is the same for both the *CW2015* ($\tau$ = 27.1 d) and the *MAST_DR23* ($\tau$ = 27.5 d).

Following Kasliwal et al. (2015), we compute the minimum timescale, $T_{Min}$, on which we observe variability in each light curve. The MAST Kepler SOC pipeline light curve suggests that $T_{Min}$ = 160 min while the *CW2015* re-processed light curve suggests that the shortest timescale on which variability may be observed is closer to $T_{Min}$ = 110 min. More significantly, the structure functions of both light curves show a smooth bowing on timescales under 10 d as compared to the mean of the ensemble of mock structure functions. This implies that the short timescale PSD slope may be a function of the timescale on which it is measured instead of being constant until some break timescale is reached, i.e. fits that are sensitive to different parts of the PSD may yield different values for the short timescale slope.

The reprocessed *CW2015* light curve of Zw 229-15, created by re-extracting the flux from a smaller aperture that more closely captures the nuclear flux as opposed to contamination from the galaxy and adjacent objects, is smoother and further from obeying the statistics of a DRW than the pipeline-processed *MAST_DR23* light curve.

## 7 COMPARISON WITH PREVIOUS STUDIES OF ZW 229-15

We compare our findings with previous variability studies of Zw 229-15. The Seyfert 1 galaxy Zw 229-15 has been extensively monitored by the ground based reverberation mapping campaign of Barth et al. (2011). Using $H\beta$ reverberation mapping over a period of 6 months, Barth et al. (2011) determined a virial black hole mass of $M_{BH} = 1.00^{+0.19}_{-0.24} \times 10^7 M_\odot$. They note that Zw 229-15 exhibits strong variability, both in the continuum as well as in the $H\beta$ line flux. Using raw SAP flux from Kepler, Mushotzky et al. (2011) performed a PSD analysis of the first 4 quarters from the light curve of Zw 229-15 and found that the short timescale PSD slope ranges between $\gamma$ = 2.96 (quarter 8) and $\gamma$ = 3.31 (quarter 6). Carini & Ryle (2012) re-processed 3 quarters of Zw 229-15 data (quarters 5 through 7) by re-defining the aperture used to compute the source flux. After stitching these quarters together, Carini & Ryle (2012) fitted the resulting light curve to 3 different PSD models (1) a simple power-law, (2) a knee model, and (3) a broken power-law. Both the knee model and the broken power law produced acceptable fits. The knee model used by Carini & Ryle (2012) is identical to the DPL model used here and was fit with short timescale slope $\gamma$ = 2.88 ± 0.21 with a turnover timescale $\tau = 92^{+27}_{-21}$ d in the rest-frame of Zw 229-15. On the other hand, the broken power law model sets the PSD slope above the break timescale to $\gamma$ = 1. Below the break timescale of $\tau = 43^{+13}_{-10}$ d, the PSD slope was found to be $\gamma$ = 2.83 ± 0.25. More recently, Williams & Carini (2015) have used the method of Carini & Ryle (2012) to fit the full *CW2015* light curve to the broken power-law model (Carini & Ryle 2012, model 3) and find short timescale slope $\gamma$ = 2.80 ± 0.43 and turnover timescale $\tau = 66.94^{+14.3}_{-11.8}$ d. While the break timescales found by Williams & Carini (2015) and Carini & Ryle (2012) are somewhat higher than those found using the structure function method of Kasliwal et al. (2015), the power law slopes are in excellent agreement.

More recently, Edelson et al. (2014) have studied the full light curve of Zw 229-15. Edelson et al. (2014) adopt a very different tactic to re-processing the Kepler light curves—instead of using a smaller aperture to attempt to isolate flux variations from the nucleus of Zw 229-15, they make the aperture much larger in an attempt to include *all* the flux from the galaxy. Consequently there is also greater contamination from neighboring sources in the sky. Their PSD analysis of the full light curve of Zw 229-15 yields a complicated two-slope PSD with long timescale slope $\gamma$ = 2.00±0.12 that transitions to a short timescale slope of $\gamma$ = 4.51 ± 0.20 at a break timescale of $\tau$ = 5.56 ± 0.95 d along with a simple power law with slope $\gamma$ = 1.28 ± 0.13 that contributes mostly around the 1 d timescale. Edelson et al. (2014) also modelled the light curve of Zw 229-15 as a Continuous-time AutoRegressive Moving Average (C-ARMA) process using the technique presented in Kelly et al. (2014) and found a bend timescale of $\tau$ = 4.00 ± 0.16 d with log-PSD slopes $\gamma = 1.99 \pm 0.01$ and $\gamma = 3.65 \pm 0.07$ on longer and shorter timescales than the bend timescale respectively. The short timescale PSD slopes found by Edelson et al. (2014) are much steeper than those inferred by us as well as those inferred by Carini & Ryle (2012). This may be partly attributable to the simplicity of the DPL PSD model that we employ—we fit a single average slope below the break timescale whereas it is evident from figure 3 that the structure function (and hence the PSD) have different slopes on different timescales below the turnover. However, variability intrinsic to neighboring sources could be a significant contaminant in the results presented in Edelson et al. (2014). We plan to apply more sophisticated C-ARMA modeling techniques to the *CW2015* version of the Zw 229-15 light curve to better characterize the observed variability (Kasliwal et al. 2015, in preparation).

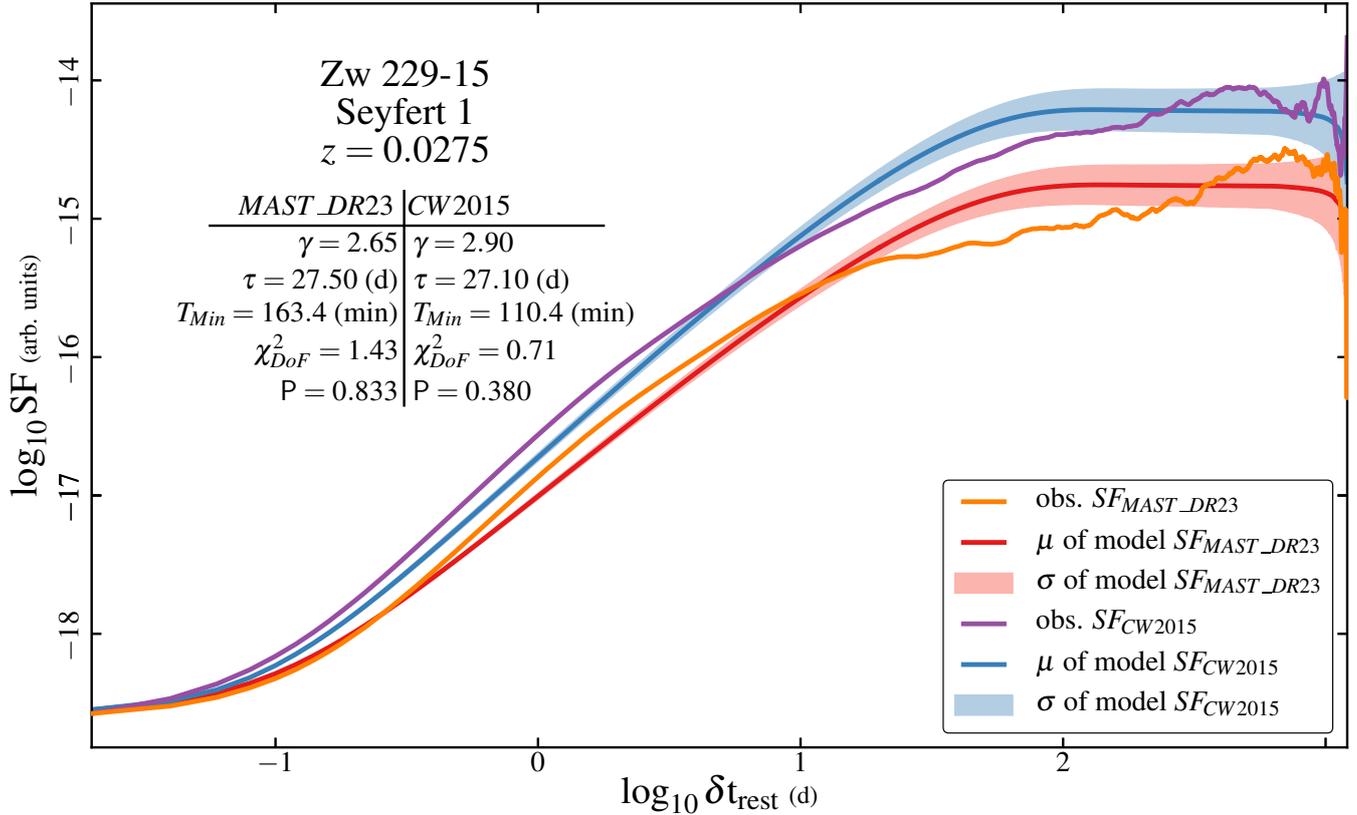

**Figure 3.** Structure Functions of the *MAST_DR23* and *CW2015* light curve of Zw 229-15.

## 8 CONCLUSIONS

We conclude that the reference Kepler MAST light curves of AGN require re-processing before inferences about the physical origins of AGN variability may be drawn from them. Such re-processing is greatly aided by the availablity of ground-based photometry to validate the optimal aperture for flux extraction. We recommend that all researchers conducting K2 campaigns (Howell et al. 2014) to monitor variable objects obtain simultaneous ground-based photometry to maximize the scientific return from the data. Although the pipeline-processed light curves may not be consistent with ground-based photometry, it is possible to re-process them to match ground-based photometry with publicly available tools. Source fluxes should be re-extracted using the smallest possible aperture to minimize the fraction of the flux from both field objects as well the AGN host galaxy, both of which are potential sources of variability given the sensitivity of the Kepler instrument. At the same time, it is very unlikely that the steep PSDs observed in the Kepler SOC pipeline AGN light curves are artificial. Re-processing the light curves will change the model inferenced from the data but is unlikely to change the overall scientific conclusion reached by Kasliwal et al. (2015)—not all AGN light curves are consistent with the DRW.

## ACKNOWLEDGEMENTS

We acknowledge support from NASA grant NNX14AL56G.

This paper includes data collected by the Kepler mission. Funding for the Kepler mission is provided by the NASA Science Mission directorate.

Some of the data presented in this paper were obtained from the Mikulski Archive for Space Telescopes (MAST). STScI is operated by the Association of Universities for Research in Astronomy, Inc., under NASA contract NAS5-26555. Support for MAST for non-HST data is provided by the NASA Office of Space Science via grant NNX09AF08G and by other grants and contracts.

**REFERENCES**

Barth A. J., et al., 2011, ApJ, 732, 121
Borucki W. J., et al., 2010, Science, 327, 977
Carini M. T., Ryle W. T., 2012, ApJ, 749, 70
Christiansen J. L., et al., 2013, Technical Report KSCI-19040-004, Kepler Data Characteristics Handbook. National Aeronautics and Space Administration, NASA Ames Research Center, Moffett Field, California
Cowperthwaite P. S., Reynolds C. S., 2014, ApJ, 791, 126
Dexter J., Agol E., 2011, ApJ, 727, L24
Edelson R., Mushotzky R., Vaughan S., Scargle J., Gandhi P., Malkan M., Baumgartner W., 2013, ApJ, 766, 16
Edelson R., Vaughan S., Malkan M., Kelly B. C., Smith K. L., Boyd P. T., Mushotzky R., 2014, ApJ, 795, 2
Emmanoulopoulos D., McHardy I. M., Uttley P., 2010, MNRAS, 404, 931
Fanelli M. N., et al., 2011, Technical Report KSCI-19081-001, Kepler Data Processing Handbook. National Aeronautics and Space Administration, NASA Ames Research Center, Moffett Field, California
Gelderman R., Guinan E., Howell S., Mattox J. R., McGruder C. H., Walter D. K., Davis D. R., Everett M., 2003, in American Astronomical Society Meeting Abstracts #202. p. 766
Gilliland R. L., 2004, Technical report, ACS CCD Gains, Full Well Depths, and Linearity up to and Beyond Saturation. Space Telescope Science Institute
Gilliland R. L., et al., 2011, ApJS, 197, 6
Goosmann R. W., Czerny B., Mouchet M., Ponti G., Dovčiak M., Karas V., Różańska A., Dumont A.-M., 2006, A&A, 454, 741
Howell S. B., et al., 2014, PASP, 126, 398
Kasliwal V. P., Vogeley M. S., Richards G. T., 2015, MNRAS, 451, 4328
Kelly B. C., Bechtold J., Siemiginowska A., 2009, ApJ, 698, 895
Kelly B. C., Becker A. C., Sobolewska M., Siemiginowska A., Uttley P., 2014, ApJ, 788, 33
Kinemuchi K., Barclay T., Fanelli M., Pepper J., Still M., Howell S. B., 2012, PASP, 124, 963
Kolodziejczak J. J., Caldwell D. A., Van Cleve J. E., Clarke B. D., Jenkins J. M., Cote M. T., Klaus T. C., Argabright V. S., 2010, in Society of Photo-Optical Instrumentation Engineers (SPIE) Conference Series. p. 1, doi:10.1117/12.857637
Kozłowski S., et al., 2010, ApJ, 708, 927
Lyubarskii Y. E., 1997, MNRAS, 292, 679
MacLeod C. L., et al., 2010, ApJ, 721, 1014
McHardy I. M., Papadakis I. E., Uttley P., Page M. J., Mason K. O., 2004, MNRAS, 348, 783
Mushotzky R. F., Edelson R., Baumgartner W., Gandhi P., 2011, ApJ, 743, L12
Peterson Bradley M., 1997, An Introduction to Active Galactic Nuclei. Cambridge University Press
Powell M. J. D., 1994, in Gomez S., Hennart J. P., eds, Mathematics and Its Applications, Vol. 275, Advances in Optimization and Numerical Analysis. Springer, Chapt. 2, pp 51–67
Revalski M., Nowak D., Wiita P. J., Wehrle A. E., Unwin S. C., 2014, ApJ, 785, 60
Rutman J., 1978, IEEE Proceedings, 66, 1048
Smith J. C., et al., 2012, PASP, 124, 1000
Stumpe M. C., et al., 2012, PASP, 124, 985
Thompson S. E., et al., 2013, Technical Report KSCI-19063-001, Kepler Data Release 23 Notes. National Aeronautics and Space Administration, NASA Ames Research Center, Moffett Field, California
Timmer J., Koenig M., 1995, A&A, 300, 707
Van Cleve J. E., Caldwell D. A., 2009, Technical Report KSCI-19033, Kepler Instrument Handbook. National Aeronautics and Space Administration, NASA Ames Research Center, Moffett Field, California
Wehrle A. E., Wiita P. J., Unwin S. C., Di Lorenzo P., Revalski M., Silano D., Sprague D., 2013, ApJ, 773, 89
Williams J., Carini M. T., 2015, in American Astronomical Society Meeting Abstracts. p. #144.56

This paper has been typeset from a TeX/LaTeX file prepared by the author.